\documentclass[twoside]{article}
\usepackage{Proc_NTSE_18}

\begin{document}
\thispagestyle{plain}
\publref{Machleidt}

\begin{center}
{\Large \bf \strut
What is wrong with our current nuclear forces?
\strut}\\
\vspace{10mm}
{\large \bf 
R. Machleidt}

\noindent{
\it Department of Physics, University of Idaho, Moscow, Idaho 83844, USA} 

\end{center}

\markboth{
R. Machleidt}
{
Wrong nuclear forces?} 

\begin{abstract}
I discuss {\it ab initio} predictions for light and intermediate-mass nuclei as well as nuclear matter. Problems and open issues are outlined and an attempt is made to relate them to specific deficiencies of the chiral two- and many-nucleon forces currently in use.
In particular, I identify the softness of the $NN$ potential (due to non-locality) as one important factor 
for the improvement of microscopic predictions. This finding is very much in tune with the recent investigation by Lu {\it et al.} (arXiv:1812.10928) where---within a simple, but realistic model---it is shown that 
proper nuclear matter saturation requires a considerable amout of non-locality in the $NN$ interaction.
\\[\baselineskip] 
\end{abstract}

One of the most fundamental aims in theoretical nuclear physics is to understand nuclear structure
and reactions in terms of the basic forces between nucleons.
In spite of intensive efforts for half a century~\cite{Neg70}, this goal has not been achieved. Why?
Microscopic nuclear structure has essentially two ingredients: quantum many-body theory (QMBT) and nuclear forces. Thus, the reason for the failure can be that either our QMBT methods are wrong or our forces are deficient---or both.
Over the past two decades, a large number of many-body approaches have been developed,
refined, and
 tested~\cite{BNV13,Hag14a,Car15,Her16}, with the result that all of them generate essentially the same predictions when applied with the same forces. Hence, QMBT seems to be under control and the failure is most likely due to  presistent problems with nuclear forces. Therefore, the focus of the rest of this paper is on nuclear interactions.

As discussed in numerous review papers~\cite{MS16,ME11,EHM09}, chiral effective field theory (EFT) is presently perceived to be
the best approach to nuclear forces since it generates the forces needed (two- and many-body forces)
on an equal footing and in a systematic way.

Consequently, a large number of applications of chiral two-nucleon forces (2NFs)
together with chiral three-nucleon forces (3NFs) [and in some cases even four-nucleon forces (4NFs)] 
have been conducted in recent years.
 These investigations include few-nucleon 
reactions~\cite{Epe02,NRQ10,Viv13,Gol14,Kal12,Nav16}, the structure of light- and medium-mass nuclei~\cite{Hag12a,Hag12b,Gez13,Her13,Som14,Heb15,Hag16,Sim17,Mor17},
infinite matter at zero temperature~\cite{HS10,Heb11,Baa13,Hag14b,Cor13,Cor14,Sam15,Dri16,Tew16,MS16,Hol17}
and finite temperature~\cite{Wel14,Wel15}, and nuclear dynamics and response functions \cite{Bac09,Bar14,Rap15,Bur16,Hol16,Bir17,Rot17}.
Although satisfactory predictions have been obtained in many cases, specific problems persist.
Among them is the problem of describing the properties of medium-mass nuclei. 
For these nuclei, typically, the predicted radii are too small~\cite{Lap16}, while binding energies turn out to be too
large~\cite{Bin14}.
This has led some groups to fit the forces directly to the properties of those medium-mass 
nuclei~\cite{Eks15}. However, the resulting $NN$ potential, which has become known as 
NNLO$_{\rm sat}$~\cite{Eks15}, reproduces $NN$ data only up to 35 MeV.
Thus, the apparent success of this potential comes, in part, 
at the expenses of a satisfactory description
of $NN$ scattering above 35 MeV, which is not an acceptable solution of the problem.
The idea of the {\it ab initio} approach is that the 2NF is fixed by two-nucleon data and the 3NF by three-nucleon data, with no further adjustments allowed. Applications in systems with $A>3$
are then true predictions.

A recent study~\cite{Sim17} has provided an indication for how to overcome the overbinding problem:
In Ref.~\cite{Heb11}, a nucleon-nucleon potential denoted by
1.8/2.0(EM) 
(which fits the $NN$ data up to 290 MeV laboratory energy)
was constructed to be extremely soft.
Together with approriate 3NFs (fit to the $^3$H binding energy and the $^4$He charge radius) it was used to calculate the ground-state properties of closed shell nuclei ranging
from $^4$He to $^{78}$Ni~\cite{Sim17}. The ground-state energies were reproduced very well, while the radii came out slightly too small. In another investigation~\cite{Mor17}, in which the same forces were applied,
the structure of the light Tin isotopes were studied, reproducing both the binding energy and the
small splitting between the lowest $J^\pi=7/2^+$ and $5/2^+$ states of $^{100}$Sn.
Moreover, in Ref.~\cite{Heb11} it had been demonstrated that the 2NF + 3NF combination used in the above-cited calculations of finite nuclei
reproduces nuclear matter saturation correctly. 
Thus, not surprisingly, there is a firm link between nuclear saturation and the ground-state
properties of medium-mass and heavy nuclei.

Althought, for reasons to be discussed below, these calculations do not provide a true solution to the radius and overbinding problem, they do give us
 a clue for  how to overcome these problems:
The 2NF has to be extremely soft, in fact, the 2NF should be such that applying it alone leads to substantial overbinding.
Then adding a repulsive density-dependent 3NF contribution makes it possible to bring about the correct
nuclear matter saturation~\cite{Heb11}. 

In theory, one may also think of other ways to explain nuclear saturation.
Namely, opposite to the above scheme, one may start from a relatively repulsive 2NF, leading to
underbinding, and then adding an attractive, density-dependent 3NF contribution.
An example for this scenario is the combination of the Argonne V18 (AV18) 2NF~\cite{WSS95} plus the Urbana IX 3NF~\cite{Pud95}.
However, the nuclear matter saturation density {\it and energy} could not be reproduced by this combination~\cite{APR98} and medium-mass
nuclei are severely underbound~\cite{Lon17}. Similar problems occur, when AV18 is combined with the 
Illinois-7 3NF~\cite{Pie08,Lon17}. So, it appears that the combination of repulsive 2NF plus attractive 3NF does not work in reality.

\begin{figure}[t]
\centering
\includegraphics[width=7.5cm]{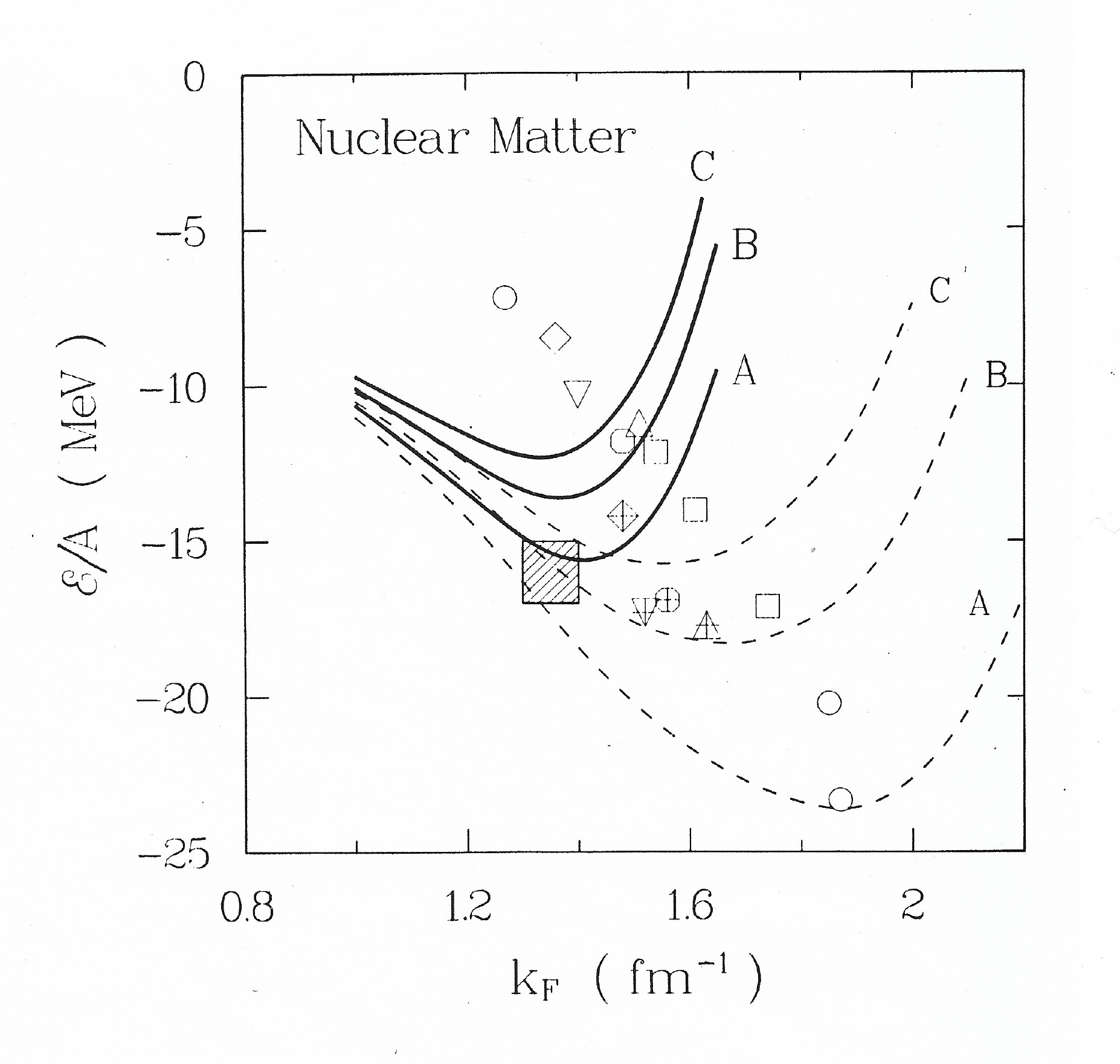}
\caption{Ground state energy per particle of symmetric nuclear matter, ${\cal E}/A$, as a function of 
the Fermi momentum, $k_F$. The dashed lines are the predictions from 2NFs while the solid lines include the 3NF effects as generated by the {\it Dirac-Brueckner-Hartee-Fock} approach. Symbols denoted the saturation points from a variety of 2NFs.
The shaded box represents the approximate empirical saturation energy and density.
From Ref.~\cite{Mac89}.}
\label{fig1}
\end{figure}

Thus, overbinding the many-body system by the 2NF and creating saturation by the 3NF contribution appears to be the only working approach. On a historical note, we mention that this is also the way how a quantitative explanation of
nuclear saturation was achieved, {\it for the first time}, applying the so-called
Dirac-Brueckner-Hartree-Fock approach~\cite{BM84,Mac89,BM90,AS03,Sam08,Sam12,MSM17}, see Fig.~\ref{fig1}.

However, the investigations of Refs.~\cite{Heb11,Sim17,Mor17} can only be perceived as test calculations,
because they are not fully consistent. The 2NF used in~\cite{Heb11,Sim17,Mor17} is very soft because it is renormalization group (RG) evolved from a harder potential. But, to preserve the attraction created 
by the softness of the potential, the induced 3NF is left out. Or, in other words, the RG evolved potential is treated like an original potential. This was useful and insightful as a test calculation to show the principle, but it cannot be viewed as a fully consistent procedure.
What we need now are fully consistent calculations, which take into account the above observations. For this, $NN$ potentials
are required
that are soft from the outset. Therefore, recently, such 
$NN$ potentials have been constructed through all order from leading-order (LO) to 
next-to-next-to-next-to-next-to-leading order
(N$^4$LO)~\cite{EMN17}.

\begin{figure}[t]
\centering
\includegraphics[width=7.0cm]{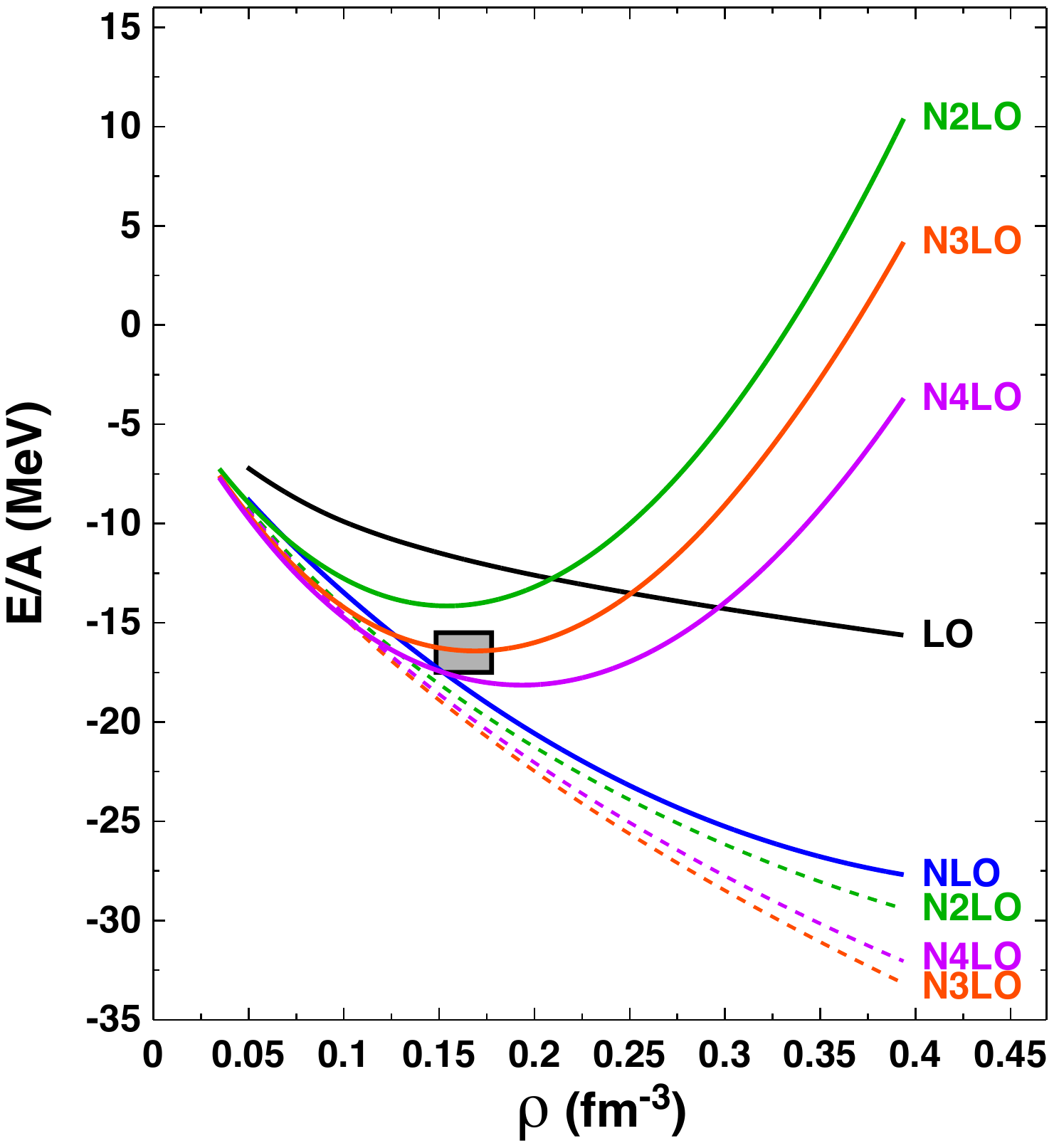}
\caption{Ground state energy per particle of symmetric nuclear matter, $E/A$, as a function of density,
$\rho$, from chiral 2NFs (dotted lines) and chiral 2NFs+3NFs (solid lines) at the denoted orders of chiral EFT.
Note that at LO and NLO, the 3NFs vanish.
The grey box represents the approximate empirical saturation energy and density.
From Ref.~\cite{Sam18}.}
\label{fig2}
\end{figure}

There are many ways to quantify the softness of a $NN$ potential.
Weinberg eigenvalues have proven to be excellent for this purpose~\cite{Hop17}.
Other, simpler parameters are  the $D$-state probability of the deuteron, $P_D$, with
low $P_D$ being a sign of softness.
The triton binding energy, $B_t$, as predicted by the 2NF alone, is also a good indicator for smoothness.
Based upon the experiences with the potentials used in Refs.~\cite{Heb11,Sim17,Mor17},
$P_D < 4.5$\% and  $B_t > 8.0$ MeV is desirable for the necessary softness of the 2NF.
The soft $NN$ potemtials of Ref.~\cite{EMN17} complemented by suitable 3NFs are generating 
promising nuclear matter predictions~\cite{DHS17,Sam18}, cf.\ Fig.~\ref{fig2}.

The softness of these potentials can be clearly attributed to their non-local charcter.
This finding is very much in tune with the recent investigation of Ref.~\cite{Lu18}  where---within a simple, but realistic model---it is shown that 
proper nuclear matter saturation requires a considerable amout of non-locality in the $NN$ interaction.

It is now of interest to apply these new interactions in
systematic studies of intermediate-mass nuclei to see if the anticipated improvements of the miscroscopic predictions do occur.
\\ \\
This research is supported in part by the US Department of Energy under Grant No.\ DE-FG02-03ER41270.

\end{document}